# Pressure induced superconducting dome in LaNiGa$_2$

Yanan Zhang[1], Dajun Su[1], Zhaoyang Shan[1], Yunshu Shi[2], Rui Li[1], Jinyu Wu[1], Zihan Yang[1], Kaixin Ye[1], Fei Zhang[3], Yanchun Li[3], Xiaodong Li[3], Chao Cao[1], Valentin Taufour[2], Lin Jiao[1], Michael Smidman[1*], and Huiqiu Yuan[1,4,5,*]

[1] *Center for Correlated Matter and School of Physics, Zhejiang University, Hangzhou 310058, China*
[2] *Department of Physics and Astronomy, University of California, Davis 95616, USA*
[3] *Beijing Synchrotron Radiation Facility, Institute of High Energy Physics, Chinese Academy of Sciences, Beijing 100049, China*
[4] *State Key Laboratory of Silicon and Advanced Semiconductor Materials, Zhejiang University, Hangzhou 310058, China*
[5] *Collaborative Innovation Center of Advanced Microstructures, Nanjing 210093, China*

LaNiGa$_2$ is a time-reversal symmetry breaking superconductor with symmetry protected band crossings, making it an ideal platform for investigating the interplay between unconventional superconductivity and electronic structure topology. Here we present a transport study of LaNiGa$_2$ under pressure. The application of pressure to LaNiGa$_2$ induces a significant enhancement of the superconducting transition temperature $T_c$ at a pressure of 7 GPa. In contrast, powder X-ray diffraction (XRD) results show no evidence of structural phase transitions up to 26.3 GPa. Moreover, the ratio of band diffusivity shows a sudden increase at around 7 GPa, suggesting possible pressure-induced changes in the electronic structure that are closely linked to the evolution of superconductivity.

## 1 Introduction

Time-reversal symmetry (TRS) breaking has been an important topic in studies of unconventional superconductors, which can be revealed by detecting small spontaneous magnetic fields in the superconducting state [1–4]. One reason for this is the detection of TRS breaking in an increasing number of weakly correlated superconductors [5–12], which is in contrast to TRS breaking in strongly correlated superconductors, where the unconventional superconductivity is evidently driven by electron-electron interactions [13–18]. While evidence for both TRS breaking and nodal superconductivity have been found in CaPtAs and LaPt$_3$P [19–21], most of these weakly correlated TRS breaking superconductors appear to have fully open superconducting gaps [22–27]. Moreover, there has been an increasing recognition of topology's role in condensed matter systems, and topological superconductivity is one of the most sought-after goals in this field [28].

LaNiGa$_2$ and LaNiC$_2$ are superconductors with orthorhombic crystal structures, where the former is centrosymmetric with $T_c$=2 K [29], while the latter is noncentrosymmetric with $T_c$=2.7 K [30]. In both materials, TRS is broken in the superconducting state [5,6,31], while probes of the superconducting gap structures reveal two-gap, nodeless superconductivity [22,23,29,32]. The low symmetry of orthorhombic structures means that if TRS is broken at $T_c$, the allowed pairing states are all non-unitary triplet states [33]. Consequently, the nodeless two-gap superconductivity was reconciled with the observed TRS breaking by the proposal of a novel internally-antisymmetric non-unitary triplet (INT) pairing state [23,34], where the pairing occurs between electrons of the same spin but on different orbitals.

Recently single crystals of LaNiGa$_2$ have been successfully synthesized and x-ray diffraction (XRD) results demonstrate that the crystal structure actually has a non-symmorphic orthorhombic space group *Cmcm*, rather than the previously reported *Cmmm* [35]. Density functional theory calculations reveal topological features in the electronic structure protected by nonsymmorphic symmetries, namely that it hosts two Dirac lines and a Dirac loop in the $k_z = \pi/c$ plane, as well as true-Dirac points that retain their degeneracy in the presence of spin-orbit coupling [35]. The degeneracy of the states at the Fermi energy suggests that topology might be relevant to TRS breaking [36]. Muon-spin relaxation experiments on single-crystal samples are in line with the proposed fully gapped non-unitary triplet superconductivity [37], while penetration depth measurements on electron irradiated crystals were interpreted within the framework of s-wave superconductivity [38]. These experimental findings indicate that the microscopic nature of the superconducting state in LaNiGa$_2$ remains an open question.

Pressure can modulate superconductivity by changing lattice constants, carrier density and bandwidth, thereby providing another perspective for studying superconductivity. For example, high-pressure transport experiments on LaNiC$_2$ have revealed an anomalous pressure dependence of $T_c$, as well as a possible high pressure correlated electronic phase that competes with superconductivity [39], which is supported by phonon-spectrum calculations [40]. Given the close similarities between the ambient pressure superconducting properties of LaNiC$_2$ and LaNiGa$_2$, it is of particular interest to examine how the superconductivity and normal state of LaNiGa$_2$ are tuned by hydrostatic pressure. In this study, we report electrical resistivity and powder XRD measurements of LaNiGa$_2$ single crystals under pressure. We find a dome-shaped $T_c$ in the pressure-temperature phase diagram, which reaches a maximum at 14.3 GPa, while the upper critical field $\mu_0 H_{c2}$ monotonically increases.

## 2 Experimental methods

Single crystals of LaNiGa$_2$ were synthesized using a Ga deficient self-flux technique, following Ref. [35]. Single crystals of LaNiGa$_2$ were polished and cut to pieces of approximate dimensions 120×80×20 μm$^3$, and were loaded into a BeCu diamond anvil cell (DAC) with a 400-μm-diameter culet. A 100-μm-thick pre-indented rhenium gasket was covered with boron nitride for electrical insulation and a 200-μm-diameter hole was drilled as the sample chamber. Daphne oil 7373 was used as the pressure transmitting medium. The DAC was loaded together with several small ruby balls for pressure determination at room temperature using the ruby fluorescence method [41]. At ambient pressure, resistivity measurements were performed using a standard four-probe method. The sample was then cut and loaded into a diamond anvil cell, and four 15 μm diameter gold wires were attached to the sample using silver epoxy paste. In this work, we obtained the absolute values of the resistivity under pressure by normalizing the resistance measured inside the pressure cell at low pressures to the resistivity obtained using the standard four-probe method. Electrical resistance measurements under pressure were performed in a Teslatron-PT system with an Oxford $^3$He refrigerator, across a temperature range of 0.3 to 300 K with a maximum applied magnetic field of 8 T. The ambient pressure resistivity measurements were performed using a Quantum Design Physical Property Measurement System (PPMS).

X-ray diffraction (XRD) measurements were carried out using beamline 4W2 of the Beijing Synchrotron Radiation Facility. Symmetric diamond anvil cells with cullet sizes of 300 μm were used in the scattering experiments with a downstream opening angle of 60°, and a monochromatic x-ray beam with a wavelength of 0.6199 Å was employed in all the measurements. The pressures were determined by the ruby fluorescence method at the sample temperature. The Fit2d program was used for image integration and the XRD patterns were fitted using the Fullprof program with the Le Bail method. Samples used in the XRD measurements were obtained by crushing single crystals.

## 3 Results and Discussion

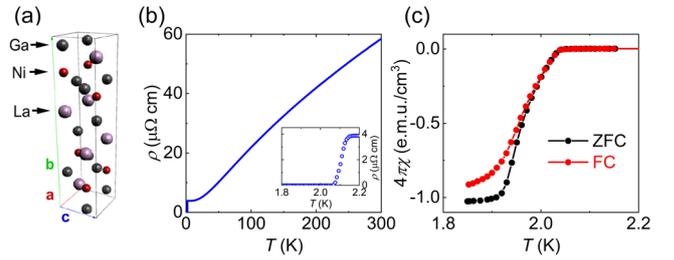

**Figure 1** (Color online) (a) Crystal structure of LaNiGa$_2$. The temperature dependence of (b) electrical resistivity $\rho(T)$ and (c) magnetic susceptibility $\chi(T)$ of LaNiGa$_2$ under ambient pressure. The inset in (b) shows an enlargement of the resistivity near $T_c$.

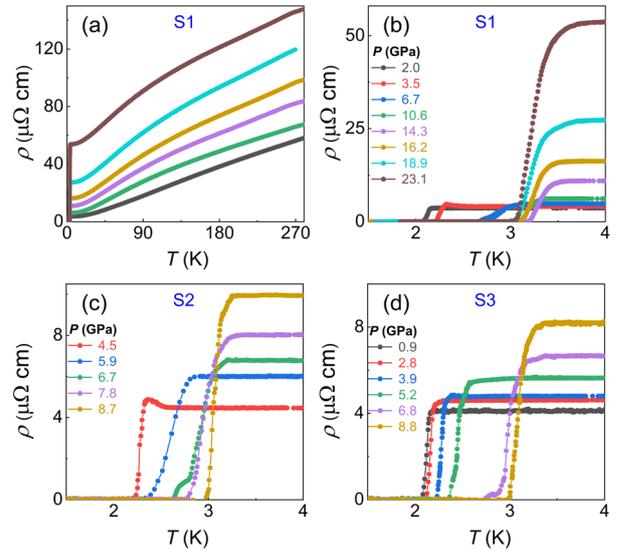

**Figure 2** (Color online) (a) Temperature-dependent resistivity $\rho(T)$ from 270 K down to 0.3 K of LaNiGa$_2$ single crystal (sample No. 1) under pressure. The low temperature $\rho(T)$ under pressure displayed for (b) sample No. 1, (c) sample No. 2, and (d) sample No. 3. Panels (a) and (b) share the same color code.

Figure 1 shows the temperature dependence of the resistivity $\rho(T)$ and magnetic susceptibility $\chi(T)$ of LaNiGa$_2$ at ambient pressure. $\rho(T)$ shows a sharp superconducting transi-tion, reaching zero at 2.1 K. The susceptibility was measured under an applied magnetic field of 1 mT after zero-field cooling (ZFC) and field cooling (FC) processes. Clear diamagnetic signals are observed in the ZFC and FC curves near the temperature where $\rho(T)$ reaches zero.

Figure 2(a) displays the $\rho(T)$ curves for sample No. 1 of LaNiGa$_2$ under pressure from 270 K to 0.3 K. The $\rho(T)$ curves consistently exhibit metallic behavior under pressure, showing a smooth decrease with decreasing temperature. This suggests the absence of pressure-induced changes to a different state, as suggested from measurements of LaNiC$_2$ [39]. When the pressure exceeds 2 GPa, a broad hump feature appears in $\rho(T)$ between 50 and 200 K, similar to the electronic crossover behavior observed in transport measurements of infinite-layer nickelate superconductors and Fe-based superconductors [42,43], which might originate from the incoherent-to-coherent crossover of the 3$d$ electrons. However, it remains an open question whether other mechanisms could also explain these observations. Figures 2(b) - 2(d) show $\rho(T)$ of three samples near $T_c$ under various pressures. Upon applying pressure, $T_c$ initially slightly increases to 2.2 K at around 4 GPa, but at higher pressures, $T_c$ increases more rapidly and reaches a maximum value of 3.2 K at around 14.3 GPa. Above 14.3 GPa, as shown in Fig. 2(b), $T_c$ monotonically decreases.

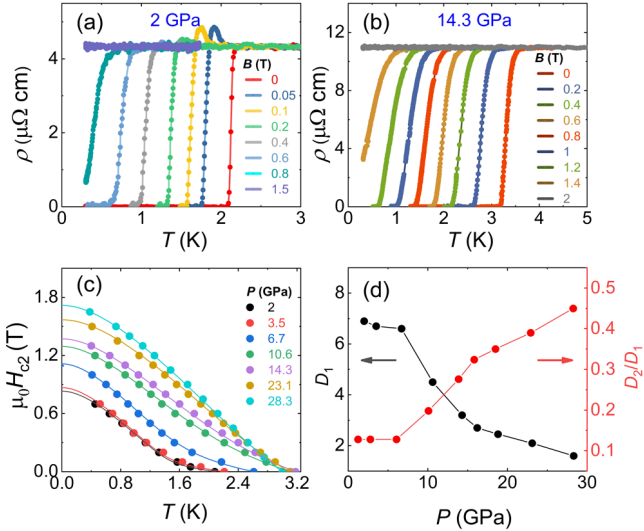

**Figure 3** (Color online) $\rho(T)$ of LaNiGa$_2$ single crystal (sample No. 1) under various applied magnetic fields at (a) 2 GPa and (b) 14.3 GPa. (c) The upper critical field $\mu_0H_{c2}(T)$ versus temperature at various pressures. The solid lines show the results from fitting using the two-band model. (d) Pressure dependence of the electronic diffusivity $D_1$ (black left axis) and the ratio $D_2/D_1$ (red right axis).

To further probe the evolution of SC in LaNiGa$_2$ under pressure, we examined the effect of applied magnetic fields along the b-axis on $T_c$. Figures 3(a) and 3(b) display the superconducting transition under different applied magnetic fields at 2 GPa and 14.3 GPa respectively, where $T_c$ is gradually suppressed by the applied field. Figure 3(c) shows the upper critical field $\mu_0H_{c2}$ (determined from the midpoint of the resistivity transitions) as a function of temperature at various pressures. At 2 GPa and 3.5 GPa, the curvature of $\mu_0H_{c2}(T)$ shows a clear upturn at 1.6 K, deviating from the single-band Werthamer-Helfand-Hohenberg (WHH) model [44]. The presence of a negative curvature in the $\mu_0H_{c2}(T)$ curve is a characteristic feature of multi-band SC. However, above 10.6 GPa, the negative curvature of $\mu_0H_{c2}(T)$ disappears, and the $\mu_0H_{c2}(T)$ more closely resembles that of single-band behavior.

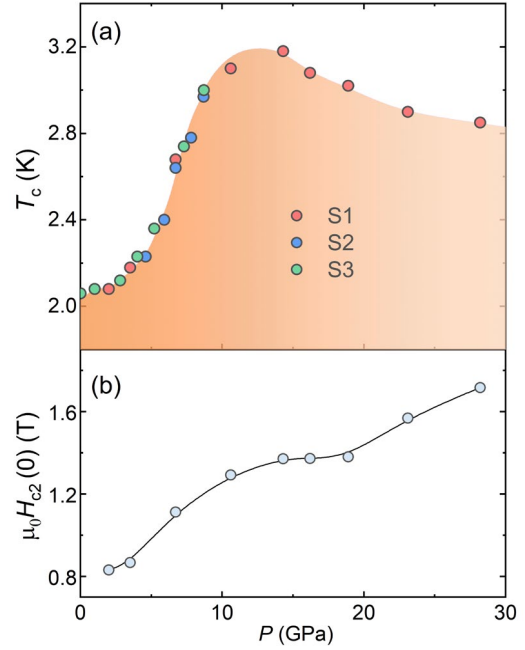

**Figure 4** (Color online) Pressure dependence of (a) the superconducting transition temperature $T_c$ for the three measured samples S1, S2, and S3, (b) the upper critical field $\mu_0H_{c2}(0)$. The shaded region in (a) is a guide for the eyes.

To better understand the evolution of $\mu_0H_{c2}(0)$ under pressure, we used a two-band model to fit the experimental data [45,46]. The two-band model takes into account both interband and intraband couplings and will be equivalent to a single-band model when the ratio of bands diffusivities $D_2/D_1 =1$. In our analysis, we use similar superconducting intraband and interband coupling constants ($\lambda_{11}$=0.156, $\lambda_{22}$=0.161, $\lambda_{12}$=$\lambda_{21}$=0.012) to those used to fit the ambient pressure superfluid density [23], so that the only free parameters are the diffusivities of the bands ($D_1$ and $D_2$). The calculated curves are shown in Fig. 3(c), where the derived zero-temperature upper critical field $\mu_0H_{c2}(0)$ shows a significant pressure dependence whereby it increases more rapidly above around 7 GPa and continues to increase even though $T_c$ decreases above 14.3 GPa, which is similar to Sr$_2$RuO$_4$ under c-axis uniaxial stress [47]. Figure 3(d) shows the evolution of the fitted $D_1$ and $D_2$ under pressure. Here $D_1$ abruptly decreases at around 7 GPa and monotonically decreases with increasing pressure up to at least 28 GPa. Correspondingly the ratio of $D_2/D_1$ increases from 0.12 at 7 GPa to 0.45 at 28.8 GPa, reflecting the shift towards sin-gle-band behavior. This suggests a possible pressure-induced change of the electronic structure. Our analysis indicates that the continuous increase of $\mu_0H_{c2}(0)$ with pressure can be accounted for by such a shift in the electronic structure while keeping the coupling constants fixed, but given the large number of model parameters, which parameters change under pressure cannot be uniquely identified.

The evolution of $T_c$ with pressure is shown in Fig. 4(a), where $T_c$ initially increases slowly with pressure, but the increase is more rapid above 7 GPa and $T_c$ reaches a maximum value of 3.2 K at around 14.3 GPa. Upon further increasing the pressure, $T_c$ monotonically decreases and reaches a value of 2.9 K at 28 GPa. The pressure dependence of the extracted $\mu_0H_{c2}(0)$ is displayed in Fig. 4(b), where $\mu_0H_{c2}(0)$ monotonically increases from 0.8 T at 2 GPa to 1.72 T at 28.3 GPa.

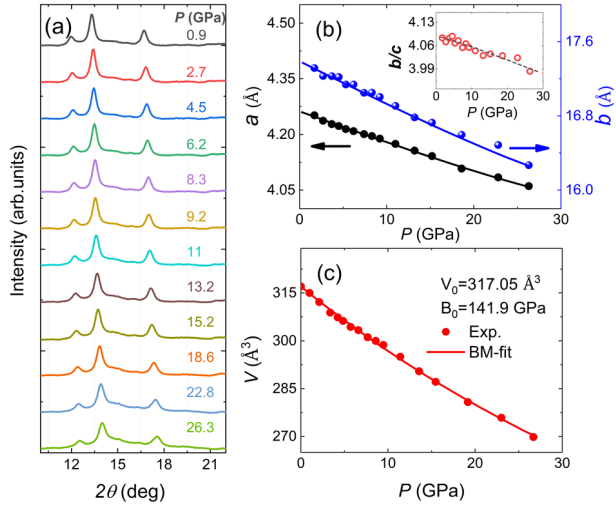

**Figure 5** (Color online) (a) XRD patterns of LaNiGa$_2$ measured under various pressures at 300 K. (b) Evolution of the lattice parameters $a$ and $b$. The inset shows the ratio $b/c$. (c) The volume of the unit cell as a function of applied pressure. The solid lines are fits to the Birch-Murnaghan equation of state.

To elucidate the origin of the enhancement and non-monotonic evolution of $T_c$ in LaNiGa$_2$, we performed XRD measurements under pressure to look for evidence of any pressure-induced structural phase transitions [48,49], as shown in Fig. 5. Our results show that the crystal structure remains unchanged with pressure up to 26.3 GPa. By analyzing the experimentally obtained XRD data using the Le Bail method [50], we extracted the lattice parameters $a$, $b$ and $c$ as a function of pressure, which are shown in Fig. 5(b). The derived unit cell volume, shown in Fig. 5(c), is well fitted with the Birch-Murnaghan equation of state [51], yielding a bulk modulus $B_0$ = 141.9 GPa. It is worth noting that $b/c$ decreases with increasing pressure, suggesting a more pronounced interlayer coupling at high pressure.

Our XRD results indicate the absence of pressure-induced structural transitions. Moreover, recent nuclear magnetic resonance (NMR) and quadrupole resonance (NQR) experiments have indicated a lack of magnetic fluctuations or enhanced paramagnetism in LaNiGa$_2$ single crystals [52], suggesting a lack of competition between superconductivity and spin fluctuations [53,54]. Therefore, the significant change of $T_c$ of LaNiGa$_2$ is likely to be associated with changes in the electronic structure, which is reflected in the suppression of the negative curvature of $\mu_0H_{c2}(T)$ as well as the deduced increase of $D_2/D_1$ under pressure. These changes coincide with the rapid increase of $T_c$ near 7 GPa, suggesting a close connection between the enhancement of $T_c$ and a possible change of electronic structure under pressure. Given the remarkable differences in the evolution of $T_c$ with pressure which reaches a maximum value, and $\mu_0H_{c2}(T)$ which monotonically increases and evolves towards single-band behavior, it is of particular interest to determine whether the superconducting order parameter also changes under pressure [55–57].

## 4 Conclusion

We have investigated the temperature-dependent resistivity of LaNiGa$_2$ at high pressures and constructed the temperature-pressure phase diagram up to 30 GPa. We find that $T_c$ increases with pressure reaching a maximum value at 14.3 GPa, and decreases at higher pressures, while $\mu_0H_{c2}(T)$ monotonically increases. However, we did not observe the evidence of pressure induced competing phases as observed in LaNiC$_2$. Moreover, powder X-ray diffraction (XRD) results show no evidence of structural phase transitions up to 26.3 GPa. Our analysis suggests that the change of $\mu_0H_{c2}(T)$ may be accounted for by the evolution of the electronic structure, but the origin of the divergence between the evolution of $T_c$ and $\mu_0H_{c2}(T)$ with pressure remains to be determined. Moreover, whether the TRS breaking superconductivity persists or if there is a change of superconducting pairing state at high pressure remains an open question. Revealing the origin of the evolution of these superconducting properties and their relation to the change of the electronic structure requires further experimental and theoretical studies.

*This work was supported by the National Key R&D Program of China (Grant No. 2022YFA1402200 and No. 2023YFA1406303), the Key R&D Program of Zhejiang Province, China (Grant No. 2021C01002), the National Natural Science Foundation of China (Grants No. 12274364, No. 12034017, No. 12174332, No. 12204408 and No. 12222410), and the Zhejiang Provincial Natural Science Foundation of China (Grant No. LR22A040002). Y.S. and V.T. acknowledge support from the UC Laboratory Fees Research Program (LFR-20-653926).*

**Conflict of interest** The authors declare that they have no conflict of interest.